\documentclass[10pt]{article}

\usepackage{graphicx}

\begin{document}

\begin{center}
{\large\bf The shape of invasion percolation clusters in random
and correlated media}
\bigskip

\author \\ Fatemeh Ebrahimi\\
{\it Department of Physics, University of Birjand, Birjand,
 Iran, 97175-615 \\}

 \end{center}

\bigskip
\begin{abstract}

The shape of two-dimensional invasion percolation clusters are
studied numerically for both non-trapping (NTIP) and trapping
(TIP) invasion percolation processes. Two different anisotropy
quantifiers, the anisotropy parameter and the asphericity are
used for probing the degree of anisotropy of clusters. We observe
that in spite of the difference in scaling properties of NTIP and
TIP, there is no difference in the values of anisotropy
quantifiers of these processes. Furthermore, we find that in
completely random media, the invasion percolation clusters are on
average slightly less isotropic than standard percolation
clusters. Introducing isotropic long-range correlations into the
media reduces the isotropy of the invasion percolation clusters.
The effect is more pronounced for the case of persisting
long-range correlations. The implication of boundary conditions
on the shape of clusters is another subject of interest. Compared
to the case of free boundary conditions, IP clusters of
conventional rectangular geometry turn out to be more isotropic.
Moreover,  we see that in conventional rectangular geometry the
NTIP clusters are more isotropic than the TIP clusters.
\end{abstract}

\bigskip

\noindent
{\bf\large I. Introduction}\\

Invasion percolation (IP)~\cite{ip1,ip2,ip3} is a dynamical
percolation process, primarily developed to describe the
evolution of the interface between two immiscible fluids in a
random porous medium. In this process, the advance of the
interface is modeled as a result of a series of discrete single
jumps of the invader (displacing fluid) into previously defender
(displacing fluid) occupied sites through the least resistant
path. The defender can be treated as an incompressible fluid.
This means that once a portion of it gets surrounded, a trap forms
and the invader cannot penetrate it further. This variant of
invasion percolation is called invasion percolation with trapping
(TIP). On the other hand, in non-trapping invasion percolation
(NTIP) which applies for compressible fluids, the invading fluid
can potentially enter any region occupied by the defender. IP has
been also used for modeling corrosion and intrusion~\cite{ip4},
simulating the melt infiltration process\cite{ip5}, and studying
random behaviour of market prices\cite{ip6}. In addition to these
applications, there are some pure scientific interests on the
subject. After all, IP is one of the simplest parameter-free
models which exhibits self-organized criticality~\cite{soc1,soc2}.

Like standard percolation~\cite{st}, invasion percolation
generates self similar fractal clusters. But unlike standard
percolation, the growth process described above, produces only a
single connected cluster. So far, much of the efforts have been
devoted on investigation of the critical
exponents~\cite{ss,sa1,sa2} and scaling properties of this
cluster~\cite{w}.  The statistics of invaded sites and the
distribution of sizes of trapped clusters in TIP have been
studied too~\cite{ip2,ip3,wb}. The shape of IP clusters has
remained an open question.

The shape of random fractals is an important physical property
that has been studied for several models including lattice
animals and percolation clusters~\cite{fam,s1,s2}, Ising
clusters~\cite{Q}, random walk~\cite{s4}, Eden clusters~\cite{s5},
bond trees~\cite{s6} and aggregates with tunable fractal
dimension~\cite{s7}. All these studies show that anisotropy is an
intrinsic property of fractal aggregates.
 Generally speaking,
the shape of a D-dimensional cluster is determined by $R^2_1 \geq
R^2_2 ... \geq R^2 _D$, where $R^2_i$'s are the eigenvalues (the
principal radii of gyration) of the cluster radius of gyration
tensor
\begin{equation}
\textbf{G}=\sum_{i=1}^N (\overrightarrow{x}_i^2\textbf{1}
-\overrightarrow{x}_i \overrightarrow{x}_i)
\end{equation}
In the above definition, $\overrightarrow{x}_i$ is the distance of
invaded site $i$ from center of mass and $N$ is the size of the
cluster. If all the $R^2_i$ are equal, the cluster is spherically
symmetric. Otherwise, it is anisotropic and we can probe the
degree of its anisotropy by defining a proper cluster anisotropy
quantifier based on the variations in the $R^2_i$~\cite{s2},
which have the following asymptotic form:
\begin{equation}
\langle R^2_i \rangle= r_i
N^{2\nu}(1+a_iN^{-\theta}+b_iN^{-1}+...)
\end{equation}
where $\nu$ is the leading scaling exponent and is equal to the
inverse of $D$, the fractal dimension of clusters. The leading
analytic correction-to-scaling term is proportional  to $N^{-1}$
and $N^{-\theta}$ represents the leading non-analytic
correction-to-scaling term. The coefficients $r_i$, $a_i$, and
$b_i$ are all independent of $N$~\cite{fam}.

Two main numerical techniques are commonly used for probing the
shape of random clusters. In the first method, proposed by Family
et al~\cite{fam}, an asymmetry measure, $A_N=R^2_D/R^2_1$, called
the anisotropy parameter of an $N$-site cluster is evaluated. The
quantity $A_N$  when properly averaged over all clusters with the
same size is denoted by $\langle A_N\rangle$ and is an estimate
of the anisotropy parameter of $N$-site clusters in the ensemble.
The case $\langle A_N\rangle=1$, corresponds to spherical
symmetry. For anisotropic objects,  $\langle A_N\rangle$ is less
than unity (the term anisotropy parameter may be misleading; the
shape of the cluster is more isotropic for larger value of
$A_N$). The asymptotic behaviour of $\langle A_\infty \rangle$ is
obtained by taking the limit $N\rightarrow \infty$. Using this
method for 2-dimension, Family et al, observed for the first time
that percolation clusters are not isotropic and estimated $\langle
A_\infty\rangle\cong0.4$ as the asymptotic value for the
anisotropy of infinitely large percolation clusters.

The method introduced by Family et al, has this advantage that
besides the shape of clusters, it provides an un-biased way of
evaluating the non-analytical correction-to-scaling
exponent~\cite{st,fam}. Nevertheless, it is difficult to treat
analytically. A more tractable approach has been suggested by
Aronovitz et al~\cite{an} and Rudnick et al~\cite{rg} based on the
definition of the asphericity $\Delta_D$ as
\begin{equation}
\Delta_D=\frac{D}{D-1} \frac{Tr\:\textbf{Q}^2}{(Tr\:\textbf{G})^2}
\end{equation}
where $\textbf{Q}=\textbf{G}-\overline{R^2}\:\textbf{I}$ and
$\overline{R^2}=D^{-1}\,Tr\: \textbf{G}$. Written in terms of
$R^2_i$ in 2-dimension, this becomes
\begin{equation}
\Delta_2=\frac{(R^2_1-R^2_2)^2}{(R^2_1+R^2_2)^2}
\end{equation}
For an isotropic cluster this quantity is equal to zero. For an
ensemble of clusters the asphericity  $\overline{\Delta}_2$ is
defined to be
\begin{equation}
\overline{\Delta}_2=\frac{\langle(R^2_1-R^2_2)^2\rangle}{\langle(R^2_1+R^2_2)^2\rangle}
\end{equation}
in which $\langle...\rangle$ denotes an ensemble average of the
quantity. Note that this quantity is different from $\langle
\Delta_2 \rangle$, the ensemble average of $\Delta_2$. Using this
method, Quandt et al~\cite{Q} obtained the value
$\overline{\Delta}_2=0.325\pm 0.006$ for the asymptotic
asphericity of two dimensional percolating clusters, showing again
that percolation clusters are not isotropic.

In this paper we study the shape of IP clusters by evaluating
both the asphericity and the anisotropy parameter. The plan of the
work is as follow. After describing the simulation method in
section II, we present the results of our extensive numerical
simulations of the  NTIP and TIP processes for completely random
media in section III. The effect of boundary conditions are
examined in section IV. Section V contains our estimations of the
shape of IP clusters when isotropic long-range correlations are
introduced into the medium. The paper is concluded at section VI.

\bigskip

\noindent {\bf\large II. Method}\\

Let us consider a  sufficiently large (effectively infinite)
square lattice with linear size $L$, and assign  to each of
lattice sites a random resistance $r$ drawn from an arbitrary
distribution $D(r)$. Starting from the center of the lattice as a
single-site invaded cluster, we follow the growth of the IP
cluster  by making a series of single jumps per time-step to the
least resistance neighbor of the cluster. Obviously, the list of
the next nearest neighbors increases rapidly with time. For the
TIP process, we should also consider the possibility of formation
of traps and discard all the trapped sites from the list of
cluster neighbors. In this work, the trapping rule has been
implemented by using the Hoshen-Kopelmn algorithm~\cite{hk}. The
search for traps is time-consuming and makes TIP simulations
much slower than NTIP simulations.

For each cluster of an arbitrary size $N$, we evaluate $R_1^1$
and $R_2^2$, the principal radii of gyration of the cluster via
diagonalization of the cluster radius of gyration tensor
$\textbf{G}$. The shape of the cluster is then characterized by
evaluating its asphericity or anisotropy parameter, as described
previously. Following the growth of the IP cluster in time, we may
calculate these values for clusters of any desired size. To
achieve highly accurate results, we estimate the mean values by
sampling the growth of IP cluster in a large number of media. The
condition of effectively infinite medium requires that none of
the IP clusters of a given size $N$ touches any boundary of the
medium. More precisely, the linear size of the lattice, $L$,
should be large enough, such that all the possible configurations
including the most anisotropic ones can potentially appear within
the lattice boundaries. Otherwise, our sampling will be biased in
favor of more isotropic clusters.

\bigskip

\noindent {\bf\large III. The shape of IP clusters in random media}\\

First we consider the shape of IP clusters in completely random
media, i.e. when $D(r)$ is chosen to be a uniform distribution.
We have followed the growth of IP clusters in $50,000$ different
samples and calculate $R^2_1$ and $R^2_2$ for a selected values of
cluster size in the range $32< N \leq 32,768$. The values of
$N$'s have chosen such that for each block of factor of two in
size (e.g. $32< N \leq 64$, $64< N \leq 128$,...,$16384<N\leq
32768$) there are $10$ equally spaced $N$'s in the logarithmic
scale. For each cluster size $N$, the anisotropy parameter
$\langle A_N \rangle$ has been calculated by averaging the ratio
$R^2_1/R^2_2$ over different samples. Then, the results have been
lumped together at the block centers. This procedure not only
helps to eliminate correction-to scaling for small
clusters~\cite{Q}, but it produces new data points which are
usually less correlated than the original data~\cite{FP}. The
same method has been applied for computing
$\langle(R^2_1-R^2_2)^2\rangle$ and
$\langle(R^2_1+R^2_2)^2\rangle$ to obtain the asphericity
parameter $\overline{\Delta}_2$ at the centre of each block. The
behaviour of anisotropy quantifiers of NTIP clusters are depicted
in fig.1 and fig.2. For comparison, the anisotropy quantifiers of
equilibrium percolation clusters are included too. These clusters
have been generated using Alexandrowicz method ~\cite{Al} which
was later modified by Grassberger~\cite{Gr}. In this method, one
starts with a single site cluster at the lattice. One of its
nearest neighbors (perimeter sites) is chosen randomly. This site
is occupied with a probability $p_c=0.592746$, the percolation
threshold of square lattice. The process continues until the
number of perimeter sites becomes zero. Only at this point, the
radius of gyration tensor is computed. We have generated
$400,000$ equilibrium percolation clusters of size $32<
N\leq32,768$ and compute the ensemble averages $\langle A_N
\rangle$, $\langle(R^2_1-R^2_2)^2\rangle$, and
$\langle(R^2_1+R^2_2)^2\rangle$ within each block.

We observe that when $N>2^{10}=1024$ the variation in all curves
becomes very small, such that for $N>2^{12}=4096$, all the curves
are effectively flat. This means the effect of
correction-to-scaling for both NTIP and percolation clusters is
negligible  and the anisotropy quantifiers have saturated. At
this limit, the anisotropy parameters of NTIP  and percolation
clusters fluctuates around $0.337\pm .001$, and $0.389\pm 0.002$,
respectively. On the other hand, the asymptotic value of the
asphericity of NTIP clusters is
$\overline{\Delta}_2=0.401\pm0.002 $, while for percolation
clusters we find $\overline{\Delta}_2=0.322 \pm 0.002$. These
observations demonstrate that NTIP clusters are less isotropic
than standard percolation clusters. This is an interesting
result, because NTIP and standard percolation clusters have the
same self-similarity dimension ($D=0.1.8959\pm 0.0001$), and
hence belong to the same universality class~\cite{ip2,sa2}.

 How are the $A_N$'s distributed? To answer
this question we have calculated $P(A)$, the normalized
distribution of $A$ for a specified cluster size say, $N=10,000$.
To this end, we divided the entire range of [0,1] to $50$ bins
with equal width $\delta=0.02$ and counted the number of clusters
with the anisotropy parameter in the range $[A - \delta, A]$. It
is seen from fig.3 that the distribution is asymmetric and quite
broad with a peak approximately located at $A \simeq 0.2 $, which
means the most probable configurations are those for them
$R_1/R_2 \approx \sqrt{0.2}=0.45$. Our calculation also shows
that the fluctuation in $A$ (not shown) is approximately equal to
$0.19$. Furthermore, we observed that the shape of $P(A)$ (and
consequently, the fluctuation) is almost independent of cluster
size $N$, if $N$ is not too small.

We have also evaluated the asphericity and the anisotropy
parameter of TIP clusters for cluster sizes in the range
$100<N<1000$. The results are presented in fig.4. As it is seen
from the figure, there is no difference in the shape of TIP and
NTIP clusters although the self-similarity dimension of these
processes differs from each other ($D=1.825 \pm 0.005$ for TIP in
square lattices ~\cite{sa2}). Both $\langle R_1^2 \rangle$ and
$\langle R_2^2 \rangle$ have the same leading exponents $2\nu$
(equation2) and hence, the anisotropy does not involve it. The
equivalence of the anisotropy quantifiers, therefore, indicates
that in addition to the value of $a_1/a_2$, the ratio of
correction-to-scaling terms is equal in these processes.
\begin{figure}
\begin{center}
\includegraphics[width= 4 in]{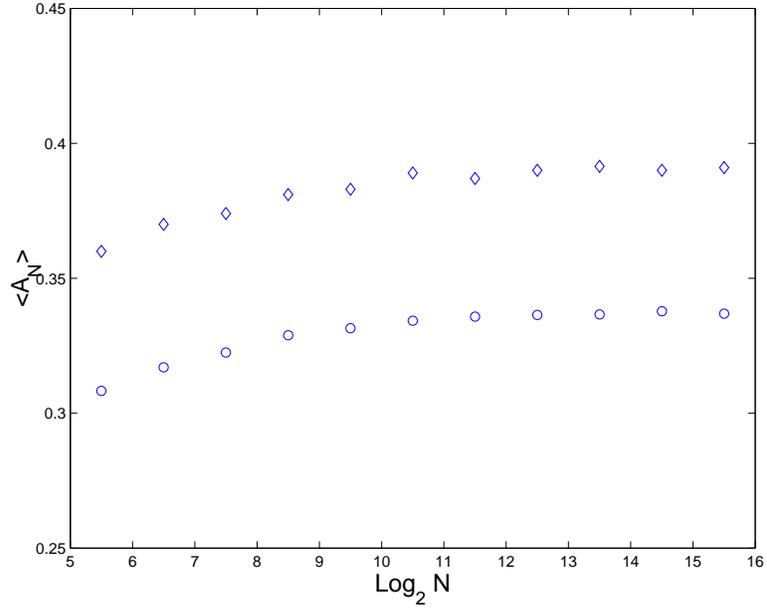}
\end{center}
\caption{Variation of $\langle A_N \rangle$, the anisotropy
parameter of equilibrium percolation clusters (diamonds) and NTIP
clusters (circles) in the range $32\leq N \leq 32,768$. For each
factor of two in size, the results have been lumped together. The
size of errorbars is smaller than the icons used. The absolute
value of error in each $\langle A_N \rangle$ is less than $0.001$
for percolation clusters. In the NTIP process, this quantity is
less than $0.0005$, since the number of samples at each block has
been much larger than the percolation case.} \label{fig.1}
\end {figure}

\begin{figure}
\begin{center}
\includegraphics[width= 4. in]{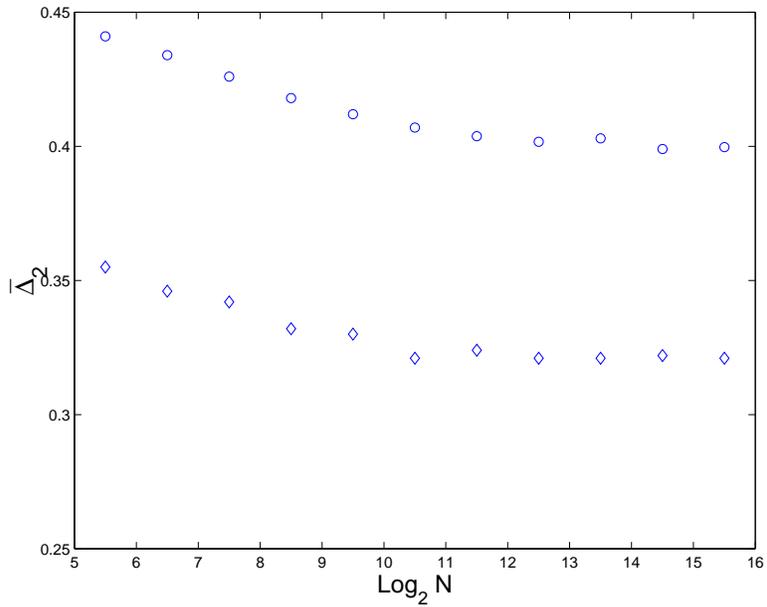}
\end{center}
\caption{Variation of $\overline{\Delta}_2$, the asphericity
parameter of equilibrium percolation clusters (diamonds) and NTIP
clusters (circles) in the range $32\leq N \leq 32,768$. The size
of errorbars is smaller than the icons used. The absolute error in
asphericity is not constant and decreases as  $N$ increases. It
is less than $0.007$ in the NTIP process and slightly greater
than $0.001$ for equilibrium percolation clusters, when $N$ is
large ($N> 4096$).} \label{fig.2}
\end {figure}

It is worth to mention that the anisotropy quantifiers are
independent of the orientation of the principal axes of the
cluster, which might be arbitrarily oriented. In fact, the
underlying ensembles of clusters are isotropic
themselves~\cite{s2}. However, isotropy of an ensemble only
implies that a given cluster conformation will appear with equal
probability in arbitrary orientations~\cite{s1}. The observed
anisotropy in the shape of clusters is a result of spontaneous
fluctuations in shape about the expected isotropic shape.  We may
relate it to the nature of the dynamics of invasion percolation.
As shown by Furuberg et al~\cite{ip3}, the advance of the
interface occurs by invading local areas in {\it bursts}; once a
new site is invaded, the interface tends to stay at that
vicinity. Quantitatively, they found that the most probable
growth after a time $t$ occurs at a distance $d_t\sim t^{1/z}$,
where $z$ is the dynamic exponent. Naturally, this local growth
might amplify any small fluctuations in the ratio of
$R_1^2/R_2^2$.

\begin{figure}
\begin{center}
\includegraphics[width= 4. in]{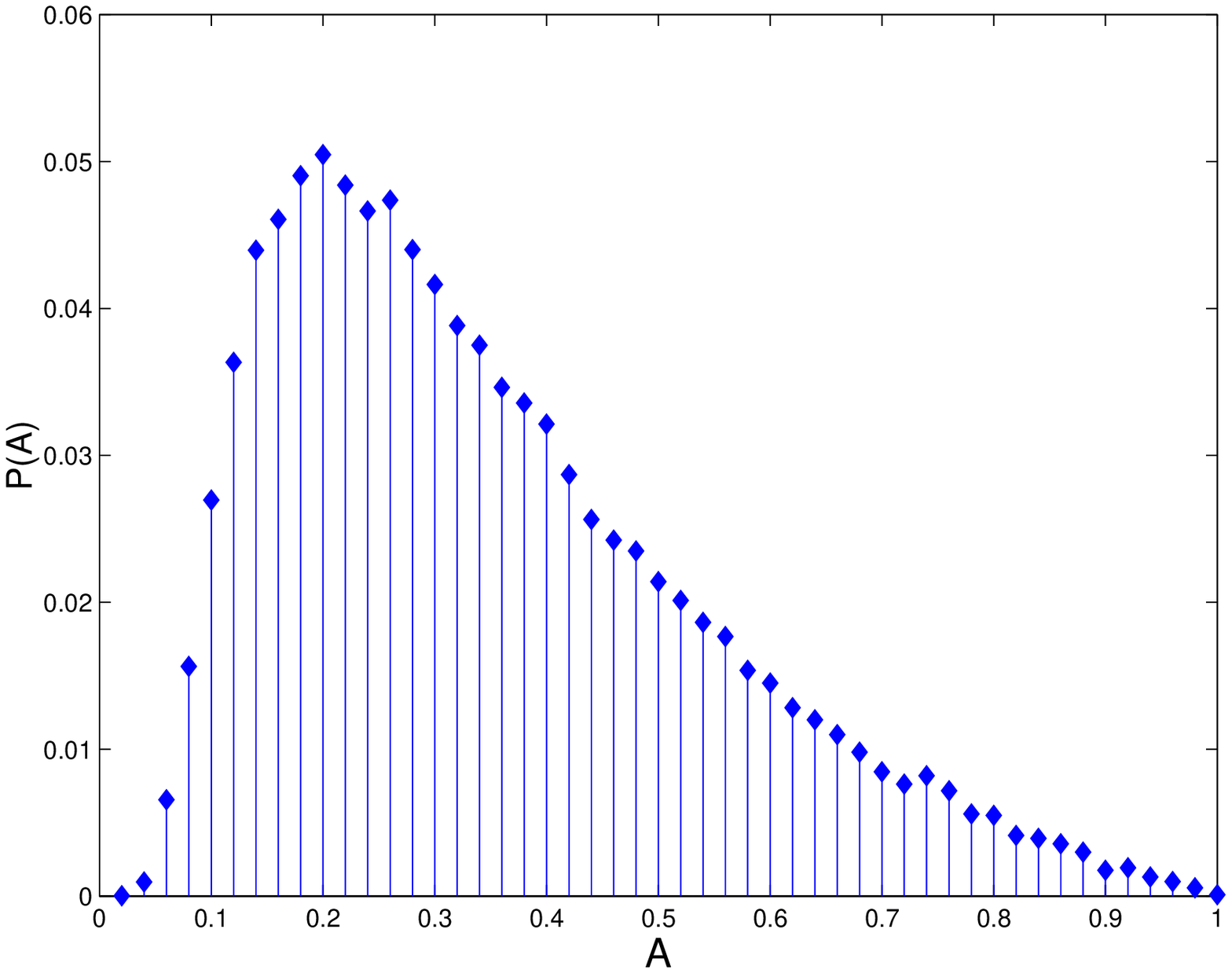}
\end{center}
\caption{$P(A)$, the normalized distribution of $A$ for NTIP
clusters of size $N=10,000$.} \label{fig.3}
\end {figure}
\begin{figure}
\begin{center}
\includegraphics[width= 4. in]{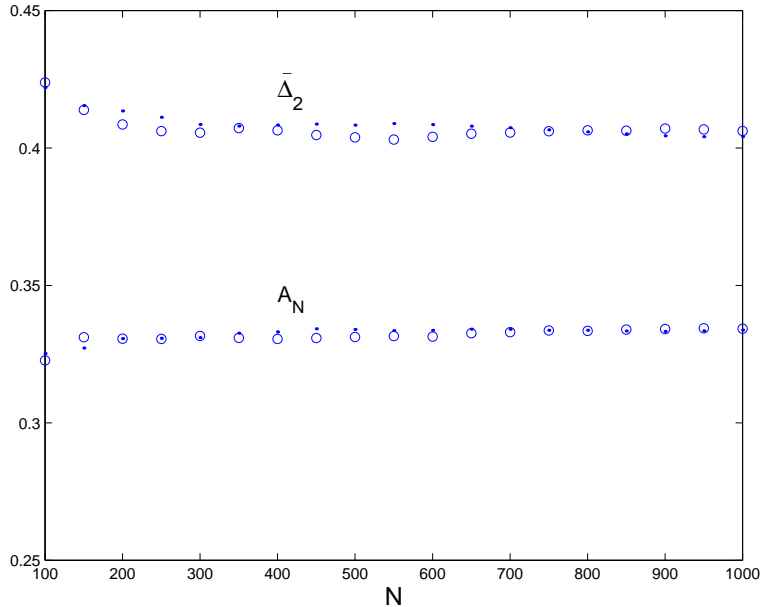}
\end{center}
\caption{Comparison between the asphericity(upper curve) and the
anisotropy parameter(lower curve) of TIP(circles) and NTIP (dots)
clusters.} \label{fig.4}
\end {figure}

\bigskip
\noindent {\bf\large IV. The shape of IP clusters in conventional geometry}\\

In the more conventional simulations of invasion percolation
processes, the host lattice is assumed to be a $L\times 2L$
rectangular lattice, and instead of the center, the invasion
process starts from one of the smaller lattice edges. The outlet
or sink is located on the opposite edge and the other two lattice
edges are assumed to be impermeable. The growth process stops at
breakthrough, when the invader reaches the outlet. In this
situation, the IP cluster connects the inlet and outlet through a
single, continuous path. The properties of this  sample spanning
cluster(SSC) within the central $L\times L$ part of the lattice,
i.e. far from inlet and outlet~\cite{ip3}, have been the subject
of intense research.

To estimate the asymptotic value of the anisotropy parameter of
the central part of SSC, we generated $10,000$ samples for each
of lattice sizes, $L=64$, $L=128$, $L=256$, and $2000$ samples of
size $L=512$. The mean anisotropy parameter $A_L$ is then
computed for each $L$. In this geometry, the mass of SSC varies in
different realizations even when $L$ is fixed. For example, in
ordinary TIP the mass of central part of SSC is $N=(5.4\pm
1.1)\times 10^4$ for $L=512$. Nevertheless, since $N$ is very
large itself, this variation does not affect the value of
$A_\infty $ via correction-to-scaling terms. In fact, our
simulations show that $A_L$ does not depend on $L$, if $L$ is
sufficiently large. The obtained value of $A_\infty$ is $0.64$
for the NTIP process, and $0.57$ for TIP process. Compared to the
previous case, the shape of SSC in both NTIP and TIP has turned
out to be more isotropic. This is because in this case, the growth
process continues even after the IP cluster touches the boundaries
of the central $L\times L$ frames. The difference between the
shape of $A_\infty$ in this geometry is a consequence of trapping
rule which limits the growth of SSC in the TIP process.

\bigskip
\noindent {\bf\large V. The effect of long range correlations on
the shape of IP clusters}\\

In many practical applications, the nature of disorder is not
completely random and there are correlations in the properties of
the medium~\cite{eur,nss2}. To investigate the effect of
correlations on the shape of IP clusters, we have considered the
case for which the distribution of the resistance of lattice sites
obeys the statistics of fractional Brownian motion (FBM)
$D_B(\textbf{x})$~\cite{fbm1,fbm2}. FBM is a stochastic process
whose increments are statistically self-similar such that its
mean square fluctuation is proportional to an arbitrary power of
the spatial displacement $\textbf{x}$

\begin{equation}
\langle [D_B(\textbf{x})-D_B(\textbf{0})]^2\rangle \sim
|\textbf{x}|^{2H}
\end{equation}
$H$ is called the Hurst exponent and determines the type of
correlations. If $H=0.5$, the above equation produces the
ordinary Brownian motion, which means that in this case there is
no correlation between different increments. If $H>0.5$, then FBM
generates positive correlations, i.e. all the points in a
neighborhood of a given point obey more or less the same trend. If
$H<0.5$, FBM is anti-persistence, i.e. a trend at a point will
not be likely followed in its immediate neighborhood.

\begin{figure}
\begin{center}
\includegraphics[width= 4. in]{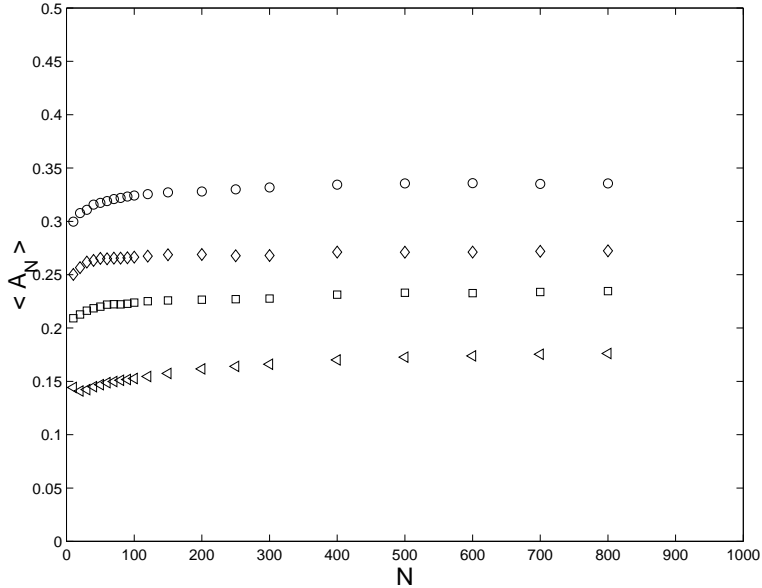}
\end{center}
\caption{The anisotropy parameters of IP clusters in media
obeying the FBM statistics; $H=0.2$ (diamonds); $H=0.5$
(squares), and $H=0.8$ (triangles). For comparison the data for
completely random media (circles) has been included too.}
\label{fig.5}
\end {figure}

The reason that we have chosen FBM process is twofold. First, FBM
generates long-range and at the same time isotropic correlations
in the field. Therefore, the host lattice retains its isotropy.
Second it has been demonstrated that such process has practical
applications in earth sciences and also reservoir engineering,
where the permeability field and also the porosity distribution
of many real oil reservoirs and aquifer follow FBM
statistic~\cite{nss2,mrs,fbm3}.

There are a number of methods which are capable of producing the
FBM statistics~\cite{mrs,fbm2}. We have used one of the most
popular one, the method of fast Fourier transformation (FFT)
filtering which is based on the fact that the power spectrum of
FBM is given by:
\begin{equation}
\mathcal{S_B}(\mathbf{\omega})=\frac{a_0}{(\omega_x^2+\omega_y^2)^p}\
\end{equation}
where $a_0$ is a numerical constant,
$\mathbf{\omega}=(\omega_1,\omega_2)$, with $\omega_i$ being the
Fourier component in the $i$th direction and $p=H+1$. In FFT
method, one starts with a white noise $W(x,y)$ defined on the
lattice sites. The power spectrum of $W(x,y)$ is constant and
independent of frequency. Therefore, filtering $W(x,y)$ with a
transfer function $\sqrt{\mathcal{S_B}(\mathbf{\omega})}$
generates another noise whose spectral density is proportional to
$\mathcal{S_B}(\mathbf{\omega})$. The method is straightforward
and fast, but it usually produces periodic noises. Therefore, one
has to produces a larger lattice and keeps only a
portion(typically $\frac{1}{4}$ in two dimensional lattices).

In fig.5 we have reported  our estimation of the anisotropy
parameters of IP clusters in media obeying the FBM statistics in
the range $10<N<800$. In this figure, we have compared the value
of $A_N$ for three different Hurst exponents, $H=0.2$
(anti-persistent correlation), $H=0.8$ (persistent correlations),
and $H=0.5$ (Brownian motion) with the results of completely
random media. The data have been obtained from averaging over
$30,000$ samples for each case. Like completely random media, we
observed no difference between the shape of NTIP and TIP clusters
(not shown). These results indicate that any deviation from
complete randomness makes the shape of invasion percolation
clusters more anisotropic. Furthermore, we find that IP clusters
in the presence of persistent correlations are less isotropic
than IP clusters of anti-persistent correlations. Based on what
has been explained in the last lines of section~III, these
effects can be assigned to the difference between dynamics of
invasion percolation in random and correlated media. In fact, we
anticipate that the burst-like growth occurs more effectively,
maybe with different dynamic exponent and amplitude (which depend
on the nature of the disorder), resulting more anisotropy in the
shape of clusters. The difference between the shape of clusters
for $H=0.2$ and $H=0.8$ is compatible with this image. The
presence of persistent long-range correlations intensify the
burst-like growth and as the result, IP clusters become more
anisotropic in this case.

\bigskip

\noindent {\bf\large VI. Conclusions}\\

The shape of clusters in IP processes have been probed numerically
by evaluating their asphericity and anisotropy parameters. The
results indicate that the shape of clusters are the same for both
TIP and NTIP processes. This conclusion does not depend on the
type of disorder in the host lattice. We found that similar to
other random fractals, generated in a variety of stochastic
processes, the invasion percolation clusters are anisotropic too.
Moreover, we observed that IP clusters are less isotropic than
standard percolation clusters. By introducing long-range
correlation into the media the clusters became more anisotropic in
shape than before. These effects might be explained according to
the dynamics of invasion percolation and the burst-like nature of
the growth process of IP clusters.

\noindent {\bf Acknowledgement}\

\noindent It is a pleasure to thank M. Sahimi who originally
pointed out the problem of the shape of invasion percolation
clusters. This work has been supported by University of Birjand
through grant $1143/1/10$.
\begin {small}

\end {small}


\begin{thebibliography}{}


\bibitem [1]{ip1} Chandler R Koplik J Lerrman K and
Willemsen J F 1982 J. Fluid Mech {\bf 119} 249.

\bibitem [2]{ip2}  Wilkinson W and  Willemsen J F 1983 J. Phys. A {\bf
16} 3365.

\bibitem [3]{ip3}  Furuberg L  Feder J  Aharony A and   J\o ssang T 1988 Phys. Rev.
Lett {\bf 61} 2117.

\bibitem [4]{ip4} Ara\'{u}jo A D  Andrade  Jr J S and
Herrmann H J 2004 Phys. Rev. E {\bf 70} 066150.

\bibitem [5]{ip5} Perham T J  Chrzan D C and De Jonghe L C 2002
Modeling Simul. Mater. Sci. Eng. {\bf 10} 103.

\bibitem [6]{ip6} Bershadskii A 2001 Physica A {\bf 300} 539.

\bibitem [7]{soc1} Stark C P 1991 Nature (London) {\bf 352} 423.

\bibitem [8]{soc2} Bak P Tang C and  Wiesenfeld K 1987 Phys. Rev. Lett. {\bf 59} 381.

\bibitem [9]{st} Stauffer D and Aharony A 1995 {\it Introduction to
Percolation Theory} (Taylor and Francis: London).

\bibitem [10] {ss}  Schwarzer S Havlin S and  Bunde A 1999 Phys. Rev. E {\bf 59} 3262.

\bibitem [11]{sa1} Sheppard A P Knackstedt M A Pinczewski W V and
Sahimi M 1999 J. Phys. A: Math. Gen. {\bf 32} L521.

\bibitem [12]{sa2}   Knackstedt M A  Sahimi S and  Sheppard A P 2002 Phys. Rev. E {\bf 65} 035101(R).

\bibitem [13]{w}  Willemsen  J F 1984 Phys. Rev. Lett. {\bf 52} 2197.

\bibitem [14]{wb}  Wilkinson D and  Barsony M 1984 J. Phys. A: Math. Gen. {\bf 17} L129.

\bibitem [15]{fam} Family F Vicsek T and Meakin P 1985 Phys. Rev. Lett {\bf 55} 641.

\bibitem [16]{s1} Aronovitz J A and Stephen M J 1987 J. Phys. A: Math. Gen. {\bf 20} 2539.

\bibitem [17]{s2} Straley J P and Stephen M J 1987 J. Phys. A: Math. Gen. {\bf 20} 6501.

\bibitem [18]{Q} Quandt S and YoungA P 1987 J. Phys. A: Math. Gen. {\bf 20} L851.

\bibitem [19]{s4}  Rudnick J Beldjenna A and  Gaspari G  1987 J. Phys. A: Math. Gen. {\bf 20} 971
; Gaspari G Rudnick J and  Beldjenna A 1987 J. Phys. A: Math.
Gen. {\bf 20} 3393.

\bibitem [20]{s5} Freche P Stauffer D  and  Stanley H E 1985 J. Phys. A:
Math. Gen. {\bf 18} L1163.

\bibitem [21]{s6} Ishinabe T 1989 J. Phys. A:
Math. Gen. {\bf 22} 4419.

\bibitem [22]{s7} Thouy R and Jullien R 1997 J. Phys. A:
Math. Gen. {\bf 30} 6725.

\bibitem [23]{an}  Aronovitz  J A and Nelson D R 1987 J. Physique
{\bf 47} 1445.

\bibitem [24]{rg}  Rudnick J  and Gaspari G 1986 J. Phys. A: Math. Gen. {\bf 19}
L191.

\bibitem [25]{hk} Hoshen J and  Kopelman R 1976 Phys. Rev. B {\bf
14} 3428.

\bibitem [26]{FP}  Flyvbjerg H and Petersen H G 1989 J. Chem. Phys. {\bf
91} 461.

\bibitem [27]{Al}  Alexandrowicz Z  1980  Phys. Lett. {\bf
80A} 284.

\bibitem [28]{Gr}  Grassberger P  1983 Math. Biosci. {\bf
62} 157.

\bibitem[29] {eur}  Vidales A M   Miranda E  Nazzarro M  Mayagoitia V  Rojas F and
Zgrablich G 1996  Europhys. Lett. {\bf 36} 259.

\bibitem [30]{nss2}  Knackstedt M A Sahimi M and
Sheppard A P  2000 Phys. Rev. E {\bf 61} 4920.


\bibitem [31]{fbm1} Mandelbrot B B 1983 {\it The Fractal Geometery of
Nature} (W. H. Freeman and Company: New York).


\bibitem [32] {mrs}Mehrabi A R Rassamdana H and Sahimi M 1997 Phys.
Rev. E {\bf 56} 712.

\bibitem [33]{fbm3} Sahimi M 1994 J. Phys. I {\bf 4} 1263.

\bibitem [34]{fbm2} Peitgen H O and  Saupe D 1988 {\it The Science of Fractal
Images} (Springer-Verlag: New York).


\end{thebibliography}
\end{document}